\documentclass[aps,prl,onecolumn,preprint,longbibliography,superscriptaddress,titlepage]{revtex4-1}
\usepackage[utf8]{inputenc}
\usepackage{amsmath}
\usepackage{mathtools}
\usepackage{amsfonts}
\usepackage{amssymb}
\usepackage{graphicx}
\usepackage{textcomp}
\usepackage{color}
\usepackage[dvipsnames]{xcolor}

\begin{document}

\title{Characterizing and Tuning Exceptional Points Using Newton Polygons}

\author{Rimika Jaiswal}
\affiliation{Undergraduate Programme, Indian Institute of Science, Bangalore 560012, India}
\author{Ayan Banerjee}
\affiliation{Solid State and Structural Chemistry Unit, Indian Institute of Science, Bangalore 560012, India}
\author{Awadhesh Narayan}
\email{awadhesh@iisc.ac.in}
\affiliation{Solid State and Structural Chemistry Unit, Indian Institute of Science, Bangalore 560012, India}
\date{\today}

\begin{abstract}
The study of non-Hermitian degeneracies -- called exceptional points -- has become an exciting frontier at the crossroads of optics, photonics, acoustics, and quantum physics. Here, we introduce the Newton polygon method as a general algebraic framework for characterizing and tuning exceptional points. Newton polygons, first described by Isaac Newton, are conventionally used in algebraic geometry, with deep roots in various topics in modern mathematics. We propose and illustrate how the Newton polygon method can enable the prediction of higher-order exceptional points, using a recently experimentally realized optical system. Using the paradigmatic Hatano-Nelson model, we demonstrate how our method can predict the presence of the non-Hermitian skin effect. As further application of our framework, we show the presence of tunable exceptional points of various orders in $PT$-symmetric one-dimensional models. We further extend our method to study exceptional points in higher number of variables and demonstrate that it can reveal rich anisotropic behaviour around such degeneracies. Our work provides an analytic recipe to understand exceptional physics.
\end{abstract}

\maketitle

\section{Introduction}
Energy non-conserving and dissipative systems are described by non-Hermitian Hamiltonians~\cite{moiseyev2011non}. Unlike their Hermitian counterparts, they are not always diagonalizable and can become defective at some unique points in their parameter space -- called exceptional points (EPs) -- where both the eigenvalues and the eigenvectors coalesce~\cite{Heiss_2012,kato2013perturbation}. Around such an EP, the complex eigenvalues lie on self-intersecting Riemann sheets. This means that upon encircling an EP once, the system does not return to its initial state, but to a different state on another Riemann sheet, manifesting in Berry phases and topological charges~\cite{heiss2016circling,el2018non,alvarez2018topological,ozdemir2019parity,science_review,ghatak2019new,ashida2020non,bergholtz2021exceptional}.

While the notion of EPs was known theoretically for several decades, their controllable realization has only been possible recently. This has led to enormous interest and, by now, EPs are ubiquitous in acoustic~\cite{shi2016accessing,acoustic2}, optical~\cite{science_review}, photonic~\cite{yin2013,zhang2016observation,ozawa2019topological,feng2014single}, mechanical~\cite{mechanical_PhysRevB.100.054109}, and condensed matter systems~\cite{midya2018non,zhang2017,ferro_PhysRevLett.121.197201,cmpnew_PhysRevB.98.035141,cmpnew_PhysRevB.99.121101}. They also appear in the study of atomic and molecular physics~\cite{molecular_PhysRevLett.103.123003}, electronics~\cite{electronic_Stehmann_2004}, superconductivity~\cite{superconductivity_PhysRevLett.99.167003,San-Jose2016,Avila2019}, and quantum phase transitions~\cite{phasetrans_PhysRevLett.99.100601}. They can lead to a variety of intriguing phenomena such as uni-directional invisibility~\cite{lin2011unidirectional,peng2014parity}, double refraction~\cite{makris2008beam}, laser mode selectivity~\cite{feng2014single,hodaei2014parity}, non-reciprocal energy transfer~\cite{xu2016topological}, non-Hermitian skin effects~\cite{alvarez2018non,yao2018edge,kunst2018biorthogonal,yokomizo2019non,lee2019anatomy,borgnia2020non,okuma2020topological,Helbig2020,Weidemann311} and interesting quantum dynamics of exciton-polaritons~\cite{gao2018chiral,gao2015observation}, to name just a few. 

While early studies focused on second order EPs (where only two eigenvectors coalesce), very recently, the focus has shifted to higher order EPs, where more than two eigenvectors coalesce~\cite{xiao2019anisotropic,xiao2020exceptional,hepth_mandal2021symmetry}. Apart from interesting fundamental physics, they show promise for several fascinating applications~\cite{nature_EnhancedSensitivity_Hodaei2017,wiersig2014enhancing}. Higher order EPs and their unconventional phase transitions have been experimentally realized in various acoustic and photonic systems~\cite{nature_EnhancedSensitivity_Hodaei2017,science_nexus,ferrohep_PhysRevB.101.144414,hep_exp}.

Here, we introduce a new algebraic framework for characterizing such non-Hermitian degeneracies using Newton polygons. These polygons were first described by Isaac Newton, in 1676, in his letters to Oldenburg and Leibniz~\cite{brieskorn2012plane}. They are conventionally used in algebraic geometry to prove the closure of fields~\cite{algebraic} and are intimately connected to Puiseux series -- a generalization of the usual power series to negative and fractional exponents~\cite{edwards2005essays,np_notes}. Furthermore, Newton polygons have deep connections to various topics in mathematics, including homotopy theory, braid groups, knot theory and algebraic number theory~\cite{brieskorn2012plane}. We develop the Newton polygon method to study EPs and illustrate its utility in predicting higher order EPs in experimentally realized systems, as well as predicting the non-Hermitian skin effect. We also present parity and time reversal ($PT$) symmetric one-dimensional models to demonstrate how this method can provide an elegant way of tuning different system parameters to obtain a higher order EP, or to choose from a spectrum of EPs of various orders. The Newton polygon method can also be naturally extended to higher number of variables. Using such an extension we show rich anisotropic behaviour around such EPs. We hope that our results stimulate further exploration of non-Hermitian degeneracies and their applications. 

\section*{The Newton Polygon Method}

We consider a system at an EP described by the Hamiltonian $H_0(t_1,t_2,...)$, where $t_1,t_2,...$ are system-dependent parameters. If a perturbation of the form $\epsilon H_1(t_1,t_2,...)$ is now added, one can write the eigenvalues of the perturbed Hamiltonian $H(\epsilon)=H_0+\epsilon H_1$ as a Puiseux series in $\epsilon$. 
\begin{equation}
    \omega(\epsilon)=\alpha_1 \epsilon^{1/N}+ \alpha_2 \epsilon^{2/N} + ...
\end{equation}

To leading order, they have the form $\omega\sim\epsilon^{1/N}$ where $N$ is the order of the EP (see Supplementary Information~\cite{supplement} for a detailed discussion). Note that in case of multiple EPs, we can write a different expansion for each EP, absorbing the zeroth order term in each of them. The Newton polygon method gives us an algorithmic way of determining the order of an EP by evaluating the power of the leading order term in the eigenvalue expansion starting from the characteristic equation. This can be done though the following steps:

\begin{enumerate}
    \item Given a characteristic equation $p(\omega,\epsilon)=\det{H-\omega\mathbb{I}}=0$, write $p(\omega,\epsilon)$ in the form $\sum_{m,n}a_{mn}(t_1,t_2,...) \omega^m \epsilon^n $.
    \item For each term of the form $a_{mn} \omega^m \epsilon^n$ in the polynomial, plot a point $(m,n)$ in $\mathbb{R}^2$. The smallest convex shape that contains all the points plotted is called the Newton polygon.
    \item  Select a segment of the Newton polygon such that all plotted points are either on, above or to the right of it.
    The negative of the slope of this line-segment gives us the lowest order dependence of $\omega$ on $\epsilon$.
\end{enumerate}

\subsection*{Predicting higher-order EP}

To illustrate our method, we consider a recently realized optical system, consisting of three coupled resonators, that exhibits a higher order EP and an unprecedented sensitivity to changes in the environment~\cite{nature_EnhancedSensitivity_Hodaei2017}. The system can be described by a remarkably simple non-Hermitian Hamiltonian

\begin{equation}
    H(\epsilon)=
    \begin{bmatrix}
    i g+\epsilon & \kappa & 0 \\
    \kappa & 0 & \kappa \\
    0 & \kappa & -i g
    \end{bmatrix},
\label{eq:nature_model}
\end{equation}

where $g$ accounts for gain and loss, $\kappa$ is the coupling between the resonators and $\epsilon$ is the external perturbation. The characteristic equation, $p(\omega,\epsilon)=0$, reads

 \begin{equation}
 -\omega^3 + \omega^2 \epsilon + (2\kappa^2-g^2)\ \omega + ig\omega\epsilon - \kappa^2 \epsilon = 0.
 \end{equation}

\begin{figure}
     \centering
     \includegraphics[width=0.95\columnwidth]{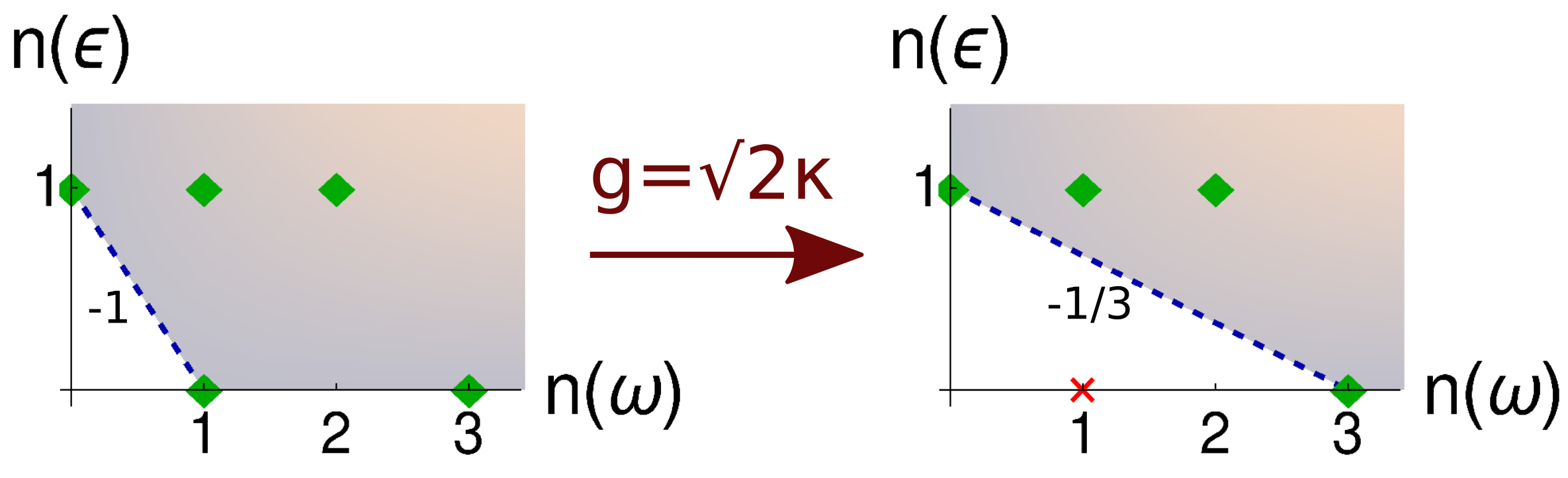}
     \caption{\textbf{Newton polygon method enabled prediction of higher order exceptional points.} Left panel shows the Newton polygon for the characteristic polynomial of $H(\epsilon)$. The line-segment which has all points either on, above, or to the right of it (shown in blue) has a slope of $-1$ implying that $\omega\sim\epsilon^1$, or, no exceptional behaviour. However, if the term corresponding to the point $(1,0)$ can be made to vanish (which occurs for $g=\sqrt{2}\kappa$), we can expect to observe exceptional behaviour. Right panel shows the Newton polygon for this case. Here, a slope of $-1/3$ implies $\omega\sim\epsilon^{1/3}$, or the presence of a third order exceptional point. Here $n(\omega$) and $n(\epsilon)$ denote the exponents of $\omega$ and $\epsilon$, respectively.}
     \label{fig:nature_np}
\end{figure}

Figure~\ref{fig:nature_np} shows the Newton polygon for $p(\omega,\epsilon)$, where each point on the graph corresponds to a term in the characteristic equation. The line-segment that contains all points on, above or to the right of it is shown in blue. This line has a slope of $-1$ implying that the lowest order dependence of the eigenvalues on $\epsilon$ has the form $\omega\sim\epsilon$. Notice, however, that if we set $g=\sqrt{2}\kappa$, the coefficient of $\omega$ vanishes and the point $(1,0)$ is no longer present in the Newton polygon. The slope of the desired line-segment is now $-1/3$ which, in turn, means that $\omega\sim\epsilon^{1/3}$, or, we have a third order EP (EP3). The Newton polygon method could thus predict the presence of an EP3 for $g=\sqrt{2}\kappa$. This is indeed what has been found in the experiments by Hodaei \textit{et al.}~\cite{nature_EnhancedSensitivity_Hodaei2017}.
 
Here, we have illustrated the use of the Newton polygon method to evaluate the degree of an EP. In addition, it also provides an algebraic way of evaluating the expansion of the eigenvalues beyond just the leading order, including the coefficients of the terms at various orders. We present a detailed discussion in the Methods sections, with worked out examples in Supplementary Information ~\cite{supplement}.

\begin{figure}
     \centering
     \includegraphics[width=\columnwidth]{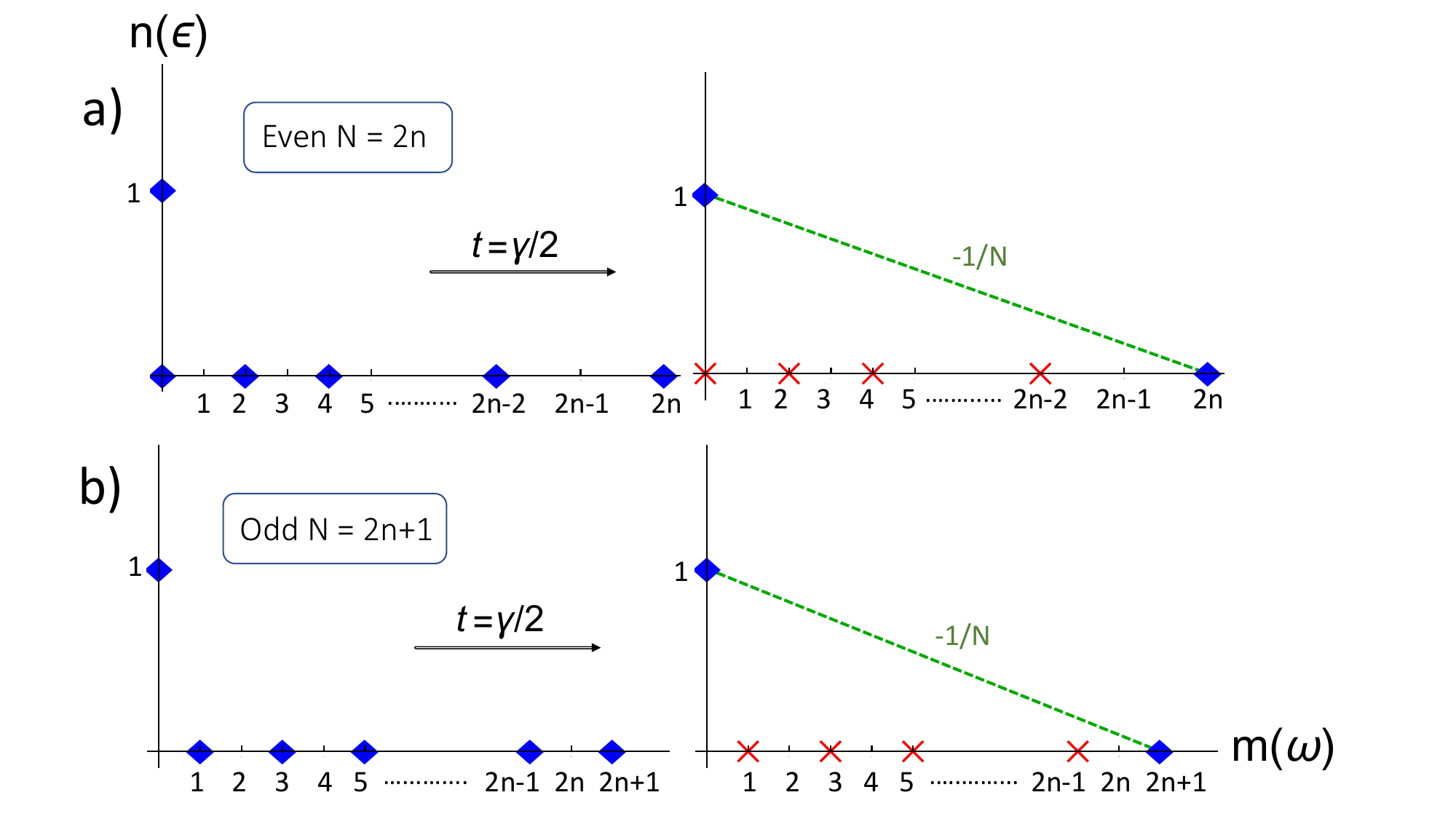}
     \caption{\textbf{Newton polygon enabled prediction of non-Hermitian skin effect.} Left panel shows the Newton polygon for the characteristic polynomial of $H_N(\epsilon)$ [Eq.~\ref{eq:HN_model_pert}]. The right panel shows the Newton Polygon for the case of $t=\gamma/2$ wherein the coefficients of all the point corresponding to $\omega^M$ vanish, other than when $M=0,N$. The line-segment which has all points either on, above, or to the right of it thus has a slope of $-1/N$, which indicates the presence of an $N$-th order EP and correspondingly the occurrence of non-Hermitian skin effect.}
     \label{fig:HN_model}
\end{figure}

\subsection*{Predicting non-Hermitian skin effect} We now use the paradigmatic Hatano-Nelson model to show that our Newton Polygon method can predict the presence of skin effect, which is a remarkable characteristic of non-Hermitian systems wherein a macroscopic number of eigenstates accumulate at the edge~\cite{alvarez2018non,yao2018edge}. The non-Hermitian skin effect is indicative of the presence of higher-order EPs as per the anomalous bulk-boundary correspondence. The Hamiltonian for the Hatano-Nelson model reads~\cite{HN_PhysRevLett.77.570}

\begin{equation}
       H=\sum_{m=1}^{N} J_R c^{\dagger}_{m+1} c_m + J_L c^{\dagger}_{m} c_{m+1},
       \label{eq:HN_model}
\end{equation}

where $J_{R/L}=t\pm \gamma/2$ are the right and left hopping amplitudes. If we now add a perturbation, $\epsilon$, coupling the first and the last sites, the Hamiltonian matrix takes the form

\begin{equation}
    H_N(\epsilon)=
    \begin{bmatrix}
    0 & t-\gamma/2 & 0 & \cdots & \epsilon \\
    t+\gamma/2 & 0 & t-\gamma/2 & \cdots & 0 \\
    0 & t+\gamma/2 & 0 & \cdots & 0\\
    \vdots & \vdots & \vdots & \ddots & \vdots \\
    \end{bmatrix}_{N \times N} .
\label{eq:HN_model_pert}
\end{equation}

The characteristic equation, in turn, is

\begin{equation}
    p(\omega,\epsilon) = (t+\gamma/2)^{N-1}\epsilon + \sum_{M=N,N-2,\cdots} z_M [t^2-(\gamma/2)^2]^{\frac{N-M}{2}} \omega^M,
\end{equation}
where each $z_M$ is a constant, $z_M\in \mathbb{Z}$. The Newton polygon for $p(\omega,\epsilon)$ is shown in the left panel of Figure~\ref{fig:HN_model}. We note that remarkably at $t=\gamma/2$, the coefficients of all the terms in $p(\omega,\epsilon)$ vanish other than the $\epsilon^1$ and the $\omega^N$ terms. The Newton polygon for this case is plotted in the right panel in Figure~\ref{fig:HN_model}. The slope of the relevant line-segment is $1/N$. This implies the presence of an $N$-th order EP and correspondingly the presence of non-Hermitian skin effect at $t=\gamma/2$, which is physically the condition for unidirectional hopping ($J_R=\gamma$, $J_L=0$). If the perturbing term $\epsilon$ is added to the bottom left element of the Hamiltonian, the other limit of purely unidirectional hopping with $J_R=0, J_L=-\gamma$ is obtained. This is also predicted from our Newton polygon approach by constructing the corresponding characteristic equation. Thus, our Newton polygon method can elegantly predict the occurrence of non-Hermitian skin effect. We note that our Newton polygon approach is able to characterize the non-Hermitian skin effect in the scenario where higher order EPs appear with an algebraic multiplicity scaling with system size while the geometric multiplicity becomes unity. In this situation, all the bulk modes may align to one state and result in the non-Hermitian skin effect.


\subsection*{Application to $PT$-symmetric 4-site model}

\begin{figure*}[tbp]
    \centering
    \includegraphics[width=0.95\textwidth]{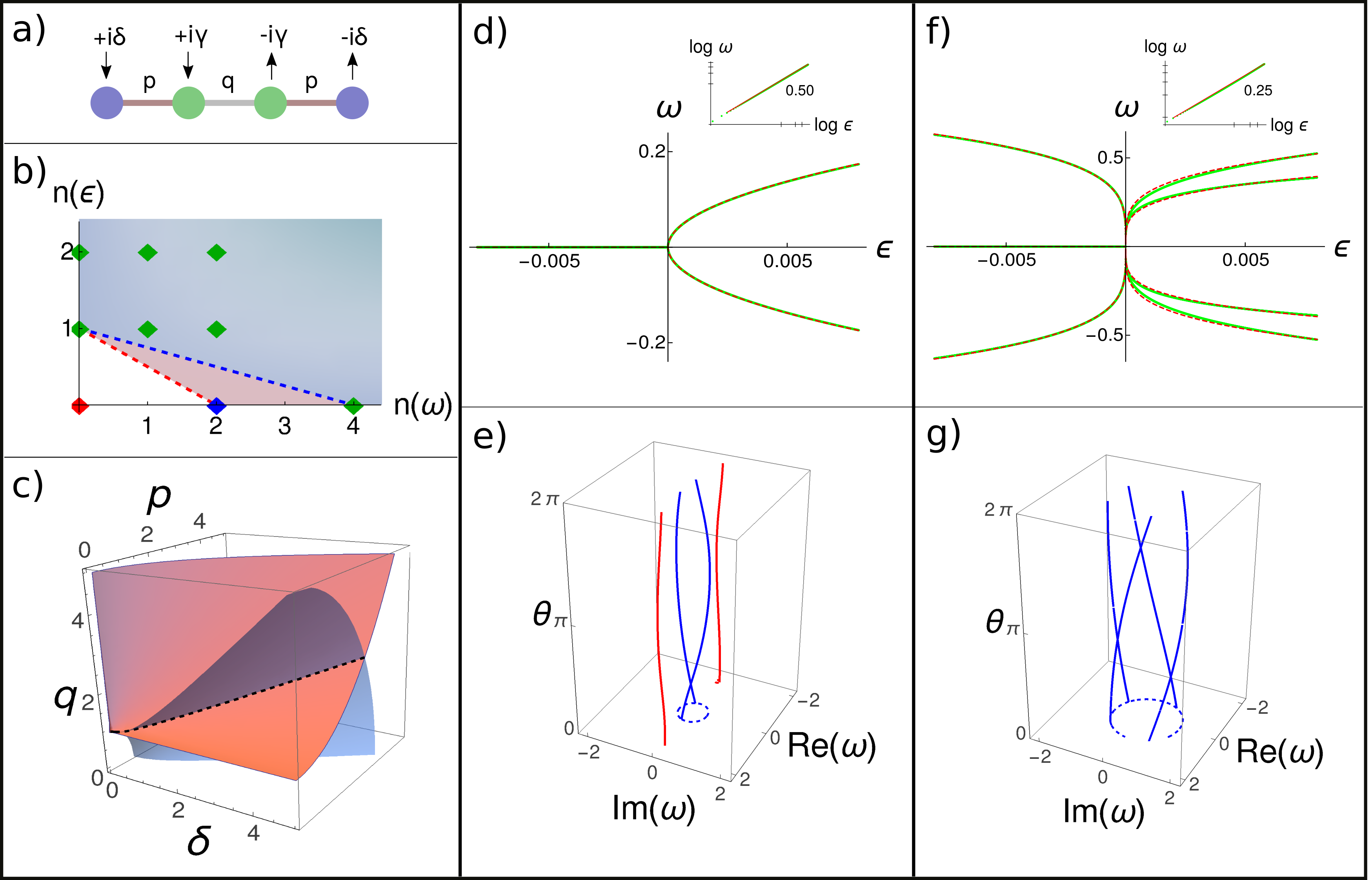}
    \caption{\textbf{Tunable higher order exceptional points in the $PT$-symmetric model via Newton polygons.} (a) Illustration of the 4-site $PT$-symmetric model which has the Hamiltonian $H_4$. (b) The Newton polygon for $H_4(\epsilon)$. The red and blue surfaces shown in (c) correspond to the regions in the parameter space where the red and blue points in the Newton polygon vanish. This predicts that one would get an EP2 if the system parameters fall on the red surface, while one would get an EP4 if they fall on the black curve (which depicts the intersection of the two surfaces). Numerical verification of these predictions are shown in middle and right panels for EP2 and EP4, respectively. (d) The numerically obtained band structure near the exceptional point EP2 (green) closely matches the predicted behaviour of $\epsilon^{1/2}$ (red). Inset shows the plot on a logarithmic scale. (f) The numerical band structure near the exceptional point EP4 (green) closely matches the predicted behaviour of $\epsilon^{1/4}$ (red). Panels (e) and (g) show the behaviour of the four bands as one encircles the exceptional points once (choosing the contour $\epsilon=0.05 \exp(i\theta)$). As expected, two of the bands swap for the middle panel (e) while the four bands undergo a cyclic permutation for the right panel (g). Parameter values for middle panel: $\delta=2,p=\sqrt{2},q=2$. Parameter values for right panel: $\delta=3,p=\sqrt{3},q=2$.}
    \label{fig:4site}
\end{figure*}

We next consider a 4-site $PT$-symmetric system as shown schematically in Figure~\ref{fig:4site}(a). We consider a general form consistent with the $PT$-symmetry. Due to parity symmetry, two hopping parameters $p$ and $q$ are sufficient to describe the couplings between the four sites. The balanced gain and loss for the outer and inner sites are given by $\delta$ and $\gamma$ respectively.
We will show that when such a system is perturbed, one can get an EP2, or an EP4 depending on the tuning of various parameters. As we shall demonstrate below, the Newton polygon method elegantly predicts the required tuning. The Hamiltonian for the 4-site system can be written as

\begin{equation}
    H_4 =
    \begin{bmatrix}
    i\delta & p & 0 & 0 \\
    p & i\gamma & q & 0 \\
    0 & q & -i\gamma & p \\
    0 & 0 & p & -i\delta \\
    \end{bmatrix}.
\end{equation}

As the overall energy scale does not affect the behaviour, we shall set $\gamma=1$ from hereon. 

If the system is perturbed, say by slightly varying one of the couplings by $\epsilon$, the perturbed Hamiltonian reads

\begin{equation}
    H_4 (\epsilon) = 
    \begin{bmatrix}
    i\delta & p+\epsilon & 0 & 0 \\
    p+\epsilon & i & q & 0 \\
    0 & q & -i & p \\
    0 & 0 & p & -i\delta \\
    \end{bmatrix}.
\end{equation}

We present the Newton polygon of the characteristic polynomial in Figure~\ref{fig:4site}(b). Observe that if the point at $(0,0)$ is absent, we would get an EP2. The part of the three-dimensional parameter space which shows a second order EP is thus the surface where the coefficient of $\omega^0\epsilon^0$ vanishes, which is shown in red in Figure~\ref{fig:4site}(c). Now, if the point $(2,0)$ was also absent, we would get an EP4. The parameter space for which this vertex vanishes is shown in blue in Figure~\ref{fig:4site}(c). The locus of EP4 is then the curve formed by the intersection of these two surface [dashed line in Figure~\ref{fig:4site}(c)]. Note that this system cannot give us an EP3 because of the $PT$-symmetry. Our Newton polygon framework, thus, enables the identification and selection of higher order EPs in a straightforward manner.

We verify these predictions from our method in several ways. First, we show that close to the EPs, our bands scale exactly as $\omega \sim \epsilon^{1/N}$, where $N$ is the order of the EP as determined from our approach [Figure~\ref{fig:4site}(d) and (f)]. The Newton polygon method also allows calculating the coefficients of the Puiseux expansions. For example, we obtain $\omega=\sqrt{\frac{2p(\delta+p^2)}{2p^2+q^2-\delta^2-1}}\epsilon^{1/2}$ to first order for EP2, which has an excellent match to the numerical fit. Derivations of such analytical expressions are presented in the Supplementary Information~\cite{supplement}. 
As a second check, we numerically show that if we encircle the EPs, the bands swap among themselves as expected. Upon encircling an EPN, $N$ eigenvalues undergo a cyclic perturbation among themselves~\cite{supplement}. For an EP2, two of the bands swap [Figure~\ref{fig:4site} (e)], while for the EP4, all four bands undergo a cyclic permutation [Figure~\ref{fig:4site} (g)].

We remark here that Zhang \textit{et al.} have studied a model for supersymmetric arrays showing an EP4~\cite{supersymmetric_PhysRevA.101.033820}. Their model is a special case of ours with $p=\sqrt{3}$, $\delta=3$, and $q=2$.

The 4-site model we presented here can show tunable second and fourth order EPs. Analogously, we have devised a 5-site $PT$-symmetric model. Using the Newton polygon framework, we have shown that it can exhibit tunable third and fifth order EPs~\cite{supplement}. Such $PT$-symmetric models have been experimentally realized in magnetic multi-layers~\cite{PTmagnetic_PhysRevB.101.144414}, waveguides~\cite{Zhong_2016,RevModPhys.88.035002}, topoelectric circuits~\cite{PTelectric_PhysRevLett.126.215302}, and photonic lattices~\cite{El-Ganainy2018,Feng2017,nanophotonics_review,anotherRev_10.1093/ptep/ptaa094}. The Newton polygon method can serve as a useful tool for tuning to EPs in these experimental platforms.

\subsection*{Extension to higher number of variables}

\begin{figure}
    \centering
    \includegraphics[width=0.6\columnwidth]{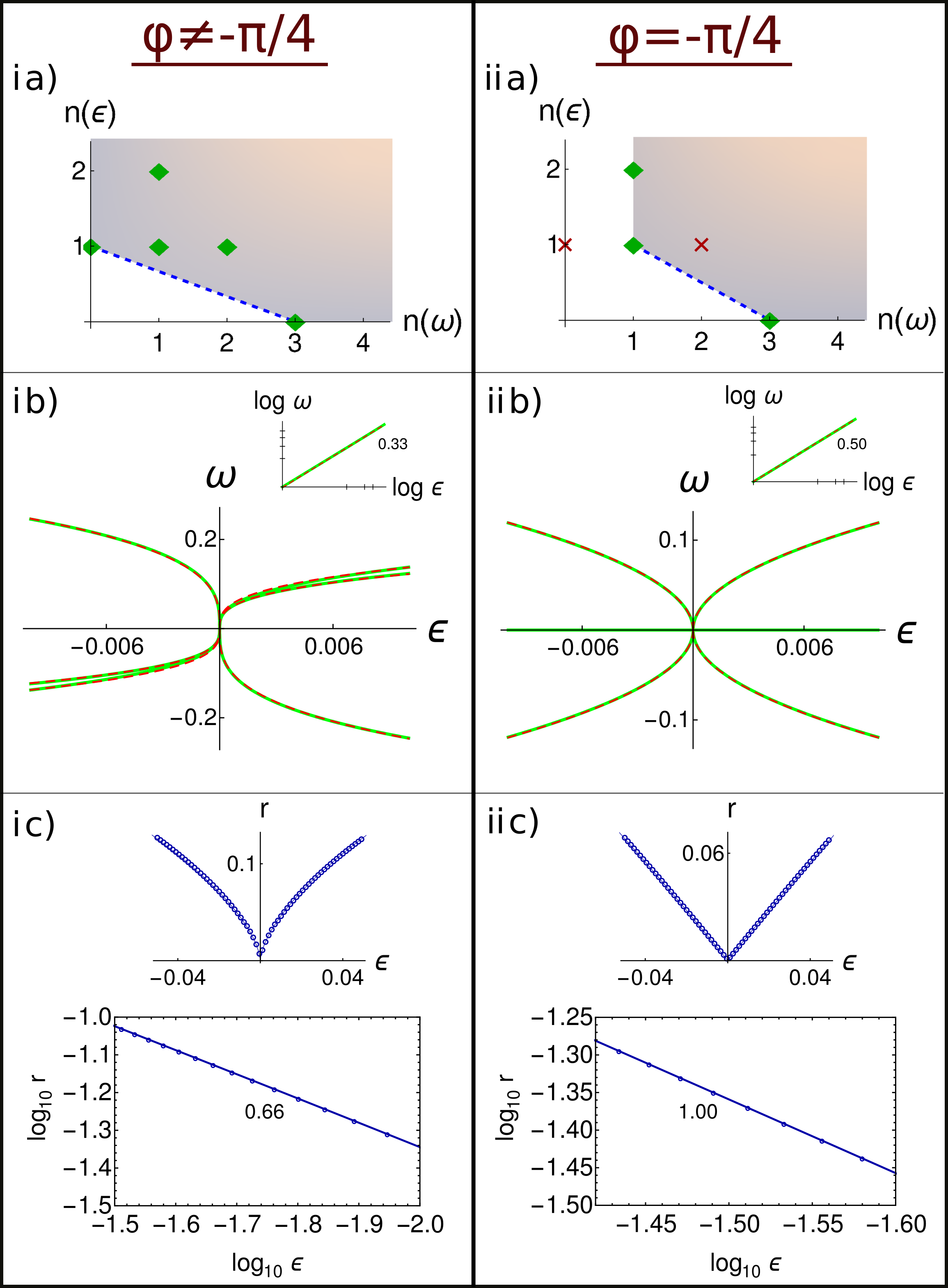}
    \caption{\textbf{Determination of anisotropic behaviour around an exceptional point using Newton polygons.} For most directions in the $\epsilon$-$\lambda$ plane making an angle $\phi$ with the $\epsilon$-axis, the Newton polygon has the form in (i.a) implying $\omega\sim\epsilon^{1/3}$.  However, the $(0,1)$ term can vanish for $\phi=-\pi/4$ giving us the Newton polygon (ii.a) which gives $\omega\sim\epsilon^{1/2}$ instead. The Newton Polygon method thus predicts second order behaviour along $\phi=-\pi/4$, but third order behaviour along all other directions.
    Numerical verification of these predictions are shown in (b) and (c). Note that the numerical band structures along a chosen direction (in green) closely match the predicted behaviours (in red) along that direction. Panels (i.c) and (ii.c) show the behaviour of the phase rigidity, $r$. The phase rigidity scales as $\epsilon^{2/3}$ for $\phi\ne-\pi/4$ as is characteristic for an EP3, while it scales as $\epsilon^{1}$ for $\phi=-\pi/4$ confirming an EP2. For plots in the left panel, $\phi=\pi/6$ was chosen as a representative direction.}
    \label{fig:twovar}
\end{figure}

The Newton polygon approach can be naturally extended to study exceptional behaviour for higher number of variables. Remarkably, it has been recently shown that if a system at an EP can be perturbed in two different ways such that $H(\epsilon,\lambda)=H_0+\epsilon H_1 + \lambda H_2$, then it possible to observe different exceptional behaviours along different directions in the $\epsilon$-$\lambda$ plane~\cite{science_nexus,exp_aniso_PhysRevLett.121.085702}. The Newton polygon framework can predict such anisotropic variations, as we show next. We consider the Hamiltonian given in Equation~\ref{eq:nature_model} and add a second perturbation, $\lambda$, to obtain

\begin{equation}
    H(\epsilon,\lambda)=
\begin{bmatrix}
i g+\epsilon & \kappa & 0 \\
\kappa & 0 & \kappa \\
0 & \kappa & -i g+\lambda
\end{bmatrix}.
\end{equation}

We can set $g=\sqrt{2}\kappa$ to obtain exceptional behaviour, as discussed earlier. If we pick any direction in the $\epsilon$-$\lambda$ plane making an angle $\phi$ with the $\epsilon$-axis, then along that direction $\lambda=\epsilon\tan{\phi}$. We can now use the Newton polygon method to determine, in one go, the order of the EPs along all directions (i.e., for all values of $\phi$).
The characteristic equation reads (for $\kappa=1$)

\begin{align}
    -\omega^3+(\tan\phi+1)\ \omega^2 \epsilon - \tan\phi\ \omega \epsilon^2 + \nonumber \\
     i\sqrt{2}(\tan\phi-1)\ \omega \epsilon  -(\tan\phi+1)\ \epsilon=0.
\end{align}

The corresponding Newton polygon is shown in Figure~\ref{fig:twovar}(a). Notice that $\omega\sim\epsilon^{1/3}$ along all directions unless $\tan\phi+1=0$ for which the point $(0,1)$ vanishes from the Newton polygon. In this case we instead obtain $\omega\sim\epsilon^{1/2}$. The Newton polygon approach thus predicts the presence of an anisotropic EP, which shows second order behaviour along $\phi=-\pi/4$, but third order behaviour along all other directions. The anisotropic nature of the EP manifests in the Newton polygon as coefficients which depend on $\phi$. We note that a similar model along two special directions has been studied before~\cite{signatures}, although a complete picture along all possible directions was lacking. Our method provides a unified way of obtaining the exceptional properties along all directions. 

We again numerically verify the predictions from the Newton polygon method in multiple ways (see Figure~\ref{fig:twovar}). We first verify our analytical expressions for the eigenvalues by fitting them to the numerical band structure. Next, we study the variation of the phase rigidity, $r$, with the perturbation. Physically, phase rigidity is a measure of the bi-orthogonality of the eigenfunctions, and helps identify EPs~\cite{muller2008exceptional,exp_aniso_PhysRevLett.121.085702,xiao2019anisotropic}. We present further details in Methods. Here, we find that the phase rigidity vanishes at the location of the EP at $\epsilon=0$ [Figure~\ref{fig:twovar} (c)]. Importantly, we observe that the phase rigidity scales as $\epsilon^{2/3}$ for $\phi\ne-\pi/4$, while it scales as $\epsilon^{1}$  for $\phi=-\pi/4$, thus confirming our predictions of the anisotropic nature of the EP.

\section*{Discussions}
We put forward a new algebraic framework using Newton polygons for characterizing EPs. Using several examples, we illustrated how the Newton polygon method enables prediction and selection of higher order EPs. Using the celebrated Hatano-Nelson model, we showed how this method allows prediction of the non-Hermitian skin effect. We also proposed an extension to higher number of variables and used it to reveal rich anisotropic behaviour around such non-Hermitian degeneracies.

Looking ahead, our analytical approach could be useful for tuning to EPs in different experimental platforms, whether it be for enhanced performance of sensors or exploring unconventional phase transitions, especially as the dimensionality and complexity of the non-Hermitian Hamiltonians in question increases.
Newton polygons play a natural role in homotopy theory, braid groups, knot theory and algebraic number theory~\cite{brieskorn2012plane,edwards2005essays}. Our work lays the foundation for exploring such connections in the context of non-Hermitian physics. With the recent growing interest in knotted and linked exceptional nodal systems~\cite{carlstrom2018exceptional,yang2019non,carlstrom2019knotted,yang2020jones,hu2021knots,zhang2021tidal,wang2020simulating}, it would be worthwhile to find a Newton polygon based characterization of such systems. In conclusion, we hope that our results inspire further exploration of EPs and their applications.

\section*{Methods}
\subsection*{Computing the coefficients and higher-order terms}
Our Newton polygon approach also provides a straightforward algorithmic method to evaluate the expansion of the eigenvalues beyond just the leading order. We describe here the steps for finding the coefficients and higher order terms in the Puiseux series expansion of eigenvalues. 

Any solution of the characteristic equation $p(\omega,\epsilon)=0$ has the form

\begin{equation}
    \omega = c_1 \epsilon^{\gamma_1} + 
    c_2 \epsilon^{\gamma_1+\gamma_2} + 
    c_3 \epsilon^{\gamma_1+\gamma_2+\gamma_3} + ... .
\end{equation}

The steps for computing $\gamma_1$ were described earlier. Once $\gamma_1$ has been determined, we can write $\omega=\epsilon^{\gamma_1} (c_1+\omega_1)$ where $\omega_1= c_2\epsilon^{\gamma_2}+ c_3\epsilon^{\gamma_2+\gamma_3}+...$. 

The steps to evaluate the other unknowns in the expansion are as follows.

\begin{enumerate}
    \item Collect the lowest order terms in $\epsilon$ in the polynomial $p(\epsilon^{\gamma_1}(c_1+\omega_1),\epsilon)$. They must cancel each other as $p(\omega,\epsilon)=0$. The first coefficient $c_1$ can be extracted from this requirement.
    \item To get the next order term, find the polynomial $p_1(\omega,\epsilon)=\epsilon^{-\beta} p(\epsilon^{\gamma_1}(c_1+\omega_1),\epsilon)$ where $\beta$ is the $y$-intercept of the line segment whose slope gave us $-\gamma_1$.
    \item Next, calculate the Newton polygon for $p_1(\omega,\epsilon)$ and repeat the steps to find $\gamma_2$ and $c_2$, and so on.
\end{enumerate}

\subsection*{Diagnosing exceptional points using phase rigidity}

The eigenvectors of non-Hermitian Hamiltonians and their characterization are strikingly different from those of their Hermitian counterparts. The system's non-Hermitian nature $(H^{\dagger}\neq H)$ suggests that the left and right eigenvectors are generally different and satisfy the following eigenvalue equations

\begin{equation}
     H |\psi \rangle = \varepsilon_i |\psi \rangle,  \quad     \langle \phi| H = \varepsilon_i \langle \phi| .
 \end{equation}

The eigenvectors form a bi-orthogonal basis consisting of both right $|\psi \rangle$ and left $\langle\phi |$ eigenvectors. They can be normalized using a bilinear product of the left and right eigenvectors, such that

\begin{equation}
   \tilde{ \psi}=\dfrac{|\psi \rangle}{\sqrt{\langle \phi|\psi\rangle}}, \quad  \tilde{ \phi}=\dfrac{\langle \phi|}{\sqrt{\langle \phi|\psi\rangle}}, \quad \langle \tilde{\phi_i}| \tilde{\psi_j}\rangle \equiv \delta_{ij},
\end{equation}

where $i, j$ correspond to distinct states. Interestingly, both the eigenvectors and eigenvalues can split and follow a directional parameter dependence while approaching an EP from different parametric directions~\cite{xiao2019anisotropic}. The phase rigidity, $r_{\alpha}$, can characterize these striking features due to extreme skewness EPs  ~\cite{rotter2009non,rotter2001dynamics,eleuch2016clustering,alvarez2018topological}. It is defined as

\begin{equation}
    r_{\alpha}= \dfrac{\langle \tilde{\phi}_{\alpha}|\tilde{\psi}_{\alpha}\rangle}{\langle \tilde{\psi}_{\alpha}|\tilde{\psi}_{\alpha}\rangle},
\end{equation}

where $\tilde{\psi}_{\alpha}$ and $\tilde{\phi}_{\alpha}$ are the normalized biorthogonal right and left eigenvectors of a state $\alpha$. The phase rigidity quantitatively measures the eigenfunctions' bi-orthogonality. At an EP the states coalesce and thus phase rigidity vanishes $(|r_{\alpha}| \rightarrow 0)$. This enables defining a critical exponent around an EP in the parameter space ~\cite{zhang2020high,tang2020exceptional}. For example, around an EP3, the scaling exponents for the phase rigidity are given by $(N-1)/N$ and $(N-1)/2$, where $N=3$ is the order of the EP. Note that different forms of the perturbation may result in different scaling of the phase rigidity. Our Newton polygon approach will allow diagnosing these EPs. The phase rigidity is also experimentally measurable, making it a very relevant quantity to look at.\\

We numerically evaluate the phase rigidity for the bivariable model with directional anisotropy, which shows that the phase rigidity scales as $\epsilon^{2/3}$ and $\epsilon^{1}$ for $\phi\neq-\pi/4$ and $\phi=-\pi/4$, respectively [see Fig. \ref{fig:twovar} (c)].

\section{Acknowledgements}
We acknowledge Madhusudan Manjunath (Department of Mathematics, Indian Institute of Technology, Bombay) and Chinmaya Kausik (Department of Mathematics, University of Michigan) for illuminating discussions.

\section{Funding}
This work was supported by Kishore Vaigyanik Protsahan Yojana (KVPY) [to R.J.]; Prime Minister's Research Fellowship (PMRF) [to A.B.]; startup grant of the Indian Institute of Science [SG/MHRD-19-0001 to A.N.] and Department of Science and Technology - Science and Engineering Research Board (DST-SERB) [SRG/2020/000153 to A.N.]

\subsection{Author Contributions}
R.J. made the connection to Newton polygons. R.J. and A.B. performed the research. R.J. wrote the manuscript with inputs from A.B. and A.N. A.N. conceived the research.

\subsection{Competing interests}
The authors declare no competing interests.

\bibliography{references}
\end{document}